*

鲍 鹏[1,2], 沈华伟[1+], 程学旗[1]

1. 中国科学院 计算技术研究所, 北京 100190
2. 北京交通大学 软件学院, 北京 100044

# Prediction of "Forwarding Whom" Behavior in Information Diffusion*

BAO Peng[1,2], SHEN Huawei[1+], CHENG Xueqi[1]

1. Institute of Computing Technology, Chinese Academy of Sciences, Beijing 100190, China
2. School of Software Engineering, Beijing Jiaotong University, Beijing 100044, China
+ Corresponding author: E-mail: shenhuawei@ict.ac.cn

**Abstract:** On online social networks, follow-ship network among users underlies the diffusion dynamics of messages; meanwhile, the structure of underlying social network determines the visibility of messages and forwarding activities in the diffusion process. Taking SinaWeibo as an example, this paper focuses on multiple exposure phenomena in information diffusion, and investigates the patterns and regularities of users' forwarding behavior among multiple exposures combined with the structure of follow-ship network. This paper analyzes the "forwarding whom" problem of users among multiple exposures in information diffusion, aiming to model and predict the forwarding behavior of individuals, combining content features, network structure, temporal and historical information. The experimental results demonstrate that the new method achieves a high accuracy of 91.3%.

**Key words:** online social network; information diffusion; multiple exposures; forwarding whom

在线社会关系网络中,用户之间的关注关系网络承载着上层的信息传播,关注关系网络的结构影响着

* The National Natural Science Foundation of China under Grant Nos. 61472400, 61232010, 61174152 (国家自然科学基金); the National Basic Research Program of China under Grant Nos. 2014CB340401, 2013CB329606 (国家重点基础研究发展计划(973计划)); the Fundamental Research Funds for the Central Universities of China under Grant No. 2015RC031 (中央高校基本科研业务费专项资金).

Received 2015-09, Accepted 2015-12.
CNKI网络优先出版: 2015-12-14, http://www.cnki.net/kcms/detail/11.5602.TP.20151214.1644.002.html

消息的可见度,并影响着信息传播过程的转发选择。以新浪微博为例,围绕信息传播中的多次暴露现象展开研究,结合用户关注关系网络的结构,探索信息传播过程中多次暴露情形下用户转发选择行为的模式和规律。针对信息传播中用户在多个暴露源下的转发选择预测问题,融合消息内容、网络结构、时序和交互历史等多方面因素,建模和预测用户转发选择。实验结果表明,新方法的预测准确率高达91.3%。

在线社会关系网络;信息传播;多次暴露;转发选择

A　　　　　　　TP391

## 1

随着信息技术的高速发展,当今人们正逐步迈进一个全新的网络化、数字化、虚拟化的工作和生活环境。越来越多的网民通过网络获取信息,并参与到信息的生成过程中,进而不断地促进网络的社会化。在用户群体规模飞速增长的同时,人们的信息需求和参与网络的方式也不断发生着变化。近年来,社交网站和社会媒体等在线社会关系网络逐渐成为互联网服务和应用的主流,典型代表包括Facebook、Twitter、微博等。这些在线社会关系网络中,人的互联和信息互联高度融合,人人参与到信息的产生与传播过程,人们信息传播和信息共享的诉求得到了极大程度的满足,并获得了前所未有的信息自主权。同时,大量的用户信息也带来了诸如信息过载、内容碎片化等问题,给学术研究和产业应用带来了新的挑战。因此,深入分析在线社会关系网络,建模和预测个体的行为具有重要意义。

在很长的一段历史时期内,由于难以获得大规模数据以提供稳定统计,信息传播和人类行为预测的研究主要局限于统计物理、社会学、认知学、心理学、行为学等学科的小规模样本分析。然而,对于小样本的研究分析结果的稳定性和代表性的质疑从未间断过。作为衔接人类社会与网络空间的纽带,在线社会关系网络汇聚了大量可感知、可计算的网络数据。这类网络数据详细记录了用户之间的网络结构以及用户产生信息的传播轨迹,这些人类活动的真实记录为研究在线社会关系网络上的信息传播以及个体行为提供了宝贵的数据资源和难得的机遇。在国内外学术界和产业界,利用在线社会关系上的在线行为数据研究人类行为逐渐成为关注的热点[1-2]。

在信息传播过程中,个人行为存在很强的随机性和自发性,不同的个体具有各异的行为模式,个体在参与信息传播时会采取不同的决策模式,使得个体传播行为呈现出差异性和不确定性。Barabási[3]发表在Nature的论文分析了人类行为的时间间隔,指出人类行为具有阵发性(burst),并给出了一种基于优先级的排队模型来解释该现象,从此拉开了人类动力学研究的序幕。Song等人[4]基于大量手机用户的通话记录,挖掘了单个用户的移动行为模式,其研究结果表明,用户的位置移动具有高达93%的可预测性。Katz[5]在社会学中提出了"两级传播"理论,强调了具有高影响力的"意见领袖"在行为传播中起着重要的作用。Wu等人[6]对Twitter全网数据进行了实证研究,并发现不同类型用户在行为周期上具有多样性。Liben-Nowell等人[7]利用大量互联网连环信的轨迹,发现连环信的传播树展现出窄而深的树状结构。Leskovec等人[8]发现,在社交网络上商品的口口相传推荐中,人与人之间的影响力会对推荐结果起到影响。Yang等人[9]分析了消息传播过程中的时序性特征,将消息传播过程聚类成6种常见类型,为理解社交网络上的用户行为提供了一些启发。Suh等人[10]在大规模社交网络数据上分析了影响个体转发行为的因素,发现消息自身的URL和Hashtag对于预测个体的转发行为具有重要指示作用。Romero等人[11]提出了粘着力(stickiness)和持久力(persistence)两个重要概念,分析不同领域内的Hashtag在Twitter上的传播过程。Myers等人[12]发现社交网络外部因素是造成信息扩散不可预测性的一个重要因素,融合了外部因素后的个体行为建模可以显著地提高信息扩散的预测准确性。Macskassy等人[13]从另外一个角度研究了Twitter上用户的转发行为,他们发现反同质性在个体转发行为上起到重要作用。Aral等人[14]认

为用户自身属性中不仅具有影响力,还有易受影响程度,进而从接受和影响两个角度对用户行为进行建模和预测。Bao等人[15-17]研究了微观结构和时序信息对个体转发行为的影响,从而更好地预测消息未来的流行度。Ugander等人[18]对信息传播的微观机制做了更深入的研究,发现个体受感染的概率不是由该个体的接触邻居个数决定的,而是由其接触邻居的连通分支个数决定的。

综上所述,在线社会关系网络中信息传播的基本规律目前尚未得到深刻理解和充分掌握,关于个体行为建模和预测的工作主要针对用户面对一条消息仅一次暴露的简单传播场景。本文将以新浪微博(http://weibo.com)为例,研究信息传播过程中的多次暴露现象(即消息在传播过程中暴露于一个用户多次),并建模和预测用户的转发选择行为。

## 2

### 2.1

微博是一种基于用户关系的信息分享、传播以及获取平台,用户可以通过Web、WAP等各种客户端组建个人社区,以不超过140字的文字更新信息,并实现即时分享。根据CNNIC关于中国社交类应用用户行为研究报告的统计(http://www.cnnic.net.cn/hlwfzyj/hlwxzbg/201408/P020140822379356612744.pdf),在2014年上半年中,43.6%的网民使用过微博,其中使用过新浪微博的网民比例最高。80.3%的新浪微博用户通过新浪微博关注新闻/热点话题,新浪微博已经成为人们了解热点信息的主要渠道之一,也是在线社会关系网络中信息传播研究的代表性场景。

本文使用的数据集是第13届在线信息系统工程会议(Web information system engineering, WISE)所发布的新浪微博数据集(http://www.wise2012.cs.ucy.ac.cy/challenge.html)。该数据集不仅包含5 800多万的用户和他们之间所形成的2亿7 000多万条关注关系,而且包括从2009年8月至2012年1月期间这些用户发布的消息及其完整的传播轨迹,其中还包括消息的主要内容属性(如是否包含嵌入式URL、相关热点事件关键词等)。

### 2.2

在新浪微博中,用户与用户之间存在着"关注"和"被关注"的关系,形成了一个关注关系网络(relationship network)。用户发出的消息(文中用户发出的消息均包括原发和转发两种消息类型)正是沿着该网络的结构被其关注者看到并传播开的。因此,关注关系网络的结构是其上信息传播的基础,不仅影响着消息的传播,同时也会受其作用而动态演化。随着网络中的连边越来越稠密,信息传播过程也会产生一些复杂的现象。

本文首先定义并探索了信息传播过程中消息的"暴露"(exposure)现象。当一条消息被用户发出后,该用户的所有关注者都会看到这条消息,则称消息暴露于该用户的关注者一次。如图1所示,一条消息在传播过程中,先后于$t_1$时刻被用户Bob转发,$t_2$时刻被用户Jim转发。在关注关系网络中,由于用户Allen同时关注了Bob和Jim两人,该消息将会两次暴露给Allen。这就是本文所研究的信息传播中的多次暴露现象。为了探索真实信息传播过程中,消息多次暴露现象是否存在,本文利用消息的传播轨迹,结合关注关系网络,统计信息传播过程中消息暴露于用户的次数分布,如图2所示。

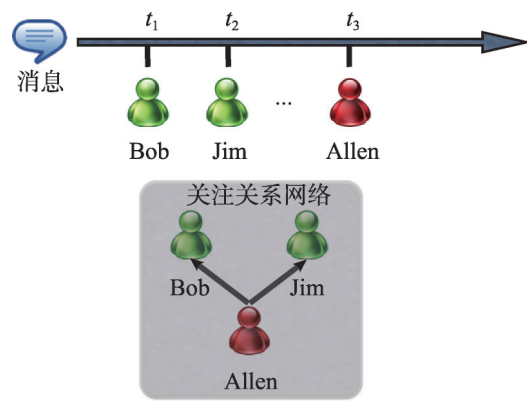

Fig.1　Multiple exposure in information diffusion
图1　信息传播中的多次暴露

## 3

本文首先提出信息传播中的用户转发选择预测问题,并对其进行形式化。然后从内容、结构、时序、

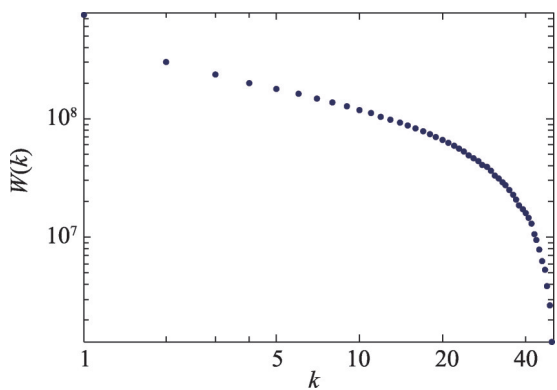

Fig.2 Statistics of multiple exposure
图2 多次暴露统计

交互历史四方面挖掘影响因素。最后建模和预测了用户的转发选择行为，并设计实验对其进行验证和分析。

### 3.1 问题描述

本文针对消息两次暴露于用户并被其转发的情形，将用户转发选择预测问题形式化成一个二分类问题。图3为用户转发选择预测问题示例，具体阐述如下。

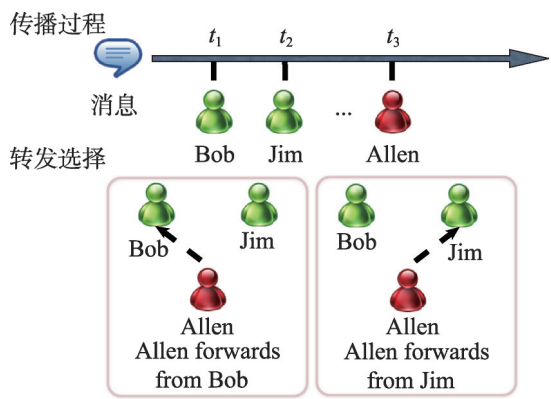

Fig.3 An example of "forwarding whom" problem
图3 "转发选择"问题示例

### 3.2 影响因素分析

#### 3.2.1 消息内容

消息内容信息对于信息传播具有重要的影响。一个直观的认识是：一条富含信息量的消息，更容易引起人们的关注和转发，而且每个人也有不同话题粒度的影响力和易受影响程度[11]。结合微博短文本内容自身的特点，已有研究结论表明，消息内容中是否包含嵌入式URL，消息是否与热点事件相关最能表达消息的内容属性，并影响用户的传播行为[15]。此外，消息当前的流行度也会对用户转发选择行为起一定的指示作用。例如对于那些被人们大量转发的热门消息，用户更愿意转发消息源或者较为权威的暴露源。

#### 3.2.2 结构特征

首先考察两个暴露源之间的结构特征。在图3中，Allen所面临的一个暴露源为Bob，另一个暴露源为Jim。在有向关注关系网络中，Bob与Jim之间存在着3种可能的关注关系：(1)Bob与Jim之间互不关注；(2)Bob与Jim之间为单向关注关系，即Bob关注Jim或者Jim关注Bob；(3)Bob与Jim之间互相关注。根据Bao等人[15]对信息传播中结构基序的研究，可以知道两个暴露源之间的关注关系是影响用户转发行为的一个重要因素。其次，入度是衡量用户影响力和可信度的一个重要因素，因此用户往往更愿意转发一个高入度的暴露源。信息传播过程中，消息的源头往往更容易受到人们的关注，因此暴露源是否为消息的原发者，是一个重要的特征。此外，由于互惠边对于信息传播也有着重要的作用[19]，暴露源是否也关注了当前用户Allen也是一个重要因素。

#### 3.2.3 时序信息

除了结构特征外，用户的转发选择行为还会受到时序信息的影响[10]。例如如果两次转发行为时间间隔过大，用户很有可能并未看到第一次暴露源，因此转发时间间隔是影响用户转发选择行为的一个因素。此外，如果一个消息是在系统不活跃时间段（如深夜）发出的，当用户第二天登录系统时，之前进行过转发的暴露源很有可能不会被用户看到，因此消息的原发时间也是一个重要因素。本文主要考察三方面因素：(1)两个暴露源之间的转发时间间隔；(2)消息传播过程中的平均转发时间间隔；(3)消息的原发时间。

#### 3.2.4 交互历史

用户之间的交互历史体现了用户之间的亲密程度以及用户一段时间内的关注兴趣，因此也会对用户的转发选择行为起到一定的影响[6,20]。本文利用当前用户是否转发过暴露源来表示用户之间的交互历史。

根据上述影响因素的分析，本文主要从消息内容、消息暴露源的结构、时序以及交互历史四方面提

取特征,用于模型的训练与预测。详细的特征选择及其描述如表1所示。

## 4

### 4.1

本文在数据集中选取2011年7月和8月的数据分别作为训练集和测试集,抽取出所有用户两次暴露于消息并最终转发的例子。最终,训练集包含10 390个例子,测试集包含11 041个例子。依据表1进行特征提取后,选择机器学习中经典的二分类模型——逻辑回归(logistic regression)模型来预测用户的转发选择行为。

文中二分类因变量 $y$ 的取值有两种可能(0和1)。以图3为例,$y=1$ 表示Allen会转发暴露源Jim,$y=0$ 表示Allen不会转发暴露源Jim。设结果 $y=1$ 的概率为 $p$,$y=0$ 的概率则为 $1-p$。假设 $x_1, x_1, \cdots, x_m$ 表示结果为 $y$ 的 $m$ 个影响因素。本文实验中 $m$ 对应表1的特征编号,用逻辑回归公式表示 $y=1$ 的概率为:

$$p = \frac{1}{1+e^{-(\beta_0+\beta_1 x_1+\beta_2 x_2+\cdots+\beta_m x_m)}}$$

式中,$\beta_0, \beta_1, \beta_2, \cdots, \beta_m$ 是模型的参数,即回归系数。

本文采用极大似然法进行回归系数的估计。假设有 $n$ 个观测样本,其观测值为 $y_1, y_2, \cdots, y_n$,其对数似然函数如下所示:

$$\ln L = \sum_{i=1}^{n}\{y_i(\beta_0+\beta_1 x_1+\beta_2 x_2+\cdots+\beta_m x_m) - \ln[1+e^{\beta_0+\beta_1 x_1+\beta_2 x_2+\cdots+\beta_m x_m}]\}$$

实验中使用准确率(Precision)、召回率(Recall)及 $F$ 值为指标对模型预测结果做出评价,其计算方法如下所示:

$$Precision = \frac{预测Allen将转发Jim且预测正确的样本总数}{预测Allen将转发Jim的样本总数}$$

$$Recall = \frac{预测Allen将转发Jim且预测正确的样本总数}{实际Allen转发Jim的样本总数}$$

Table 1　Features list

表1

| 特征类型 | 编号 | 特征描述 | 取值 |
| --- | --- | --- | --- |
| 内容特征 | 1 | 消息是否包含嵌入式URL | 1表示"是",0表示"否" |
| | 2 | 消息是否与热点事件相关 | 1表示"是",0表示"否" |
| | 3 | 消息当前流行度（用消息被转发的次数来度量） | 0表示流行度介于0~10<br>1表示流行度介于10~100<br>2表示流行度介于100~1 000<br>3表示流行度介于1 000~10 000<br>4表示流行度大于10 000 |
| 结构特征 | 4 | Bob是否关注了Jim | 1表示"是",0表示"否" |
| | 5 | Jim是否关注了Bob | 1表示"是",0表示"否" |
| | 6 | Bob是否关注了Allen | 1表示"是",0表示"否" |
| | 7 | Jim是否关注了Allen | 1表示"是",0表示"否" |
| | 8 | Jim的入度是否大于Bob | 1表示"是",0表示"否" |
| | 9 | Jim的入度是否大于Allen | 1表示"是",0表示"否" |
| | 10 | Bob的入度是否大于Allen | 1表示"是",0表示"否" |
| | 11 | Bob是否为消息的原发者 | 1表示"是",0表示"否" |
| 时序特征 | 12 | Bob和Jim转发时间间隔(小时) | 实数 |
| | 13 | 之前连续转发的平均时间差(小时) | 实数 |
| | 14 | 消息原发时间是否在活跃时间段(10 am-10 pm) | 1表示"是",0表示"否" |
| 交互历史 | 15 | Allen是否曾转发过Bob发出的消息 | 1表示"是",0表示"否" |
| | 16 | Allen是否曾转发过Jim发出的消息 | 1表示"是",0表示"否" |

$$F值 = \frac{2 \times Precision \times Recall}{Precision + Recall}$$

### 4.2

本节重点考察模型的预测性能以及各类特征对于模型预测能力的重要性,实验结果如表2所示。可以发现在融合本文所提取的四方面特征后,模型的预测准确率高达91.3%,远远高于随机猜测的50%,从而很好地预测了用户的转发选择行为。

Table 2　Experimental results
表2

| 方法 | Precision | Recall | F值 |
|---|---|---|---|
| Our method | **0.913** | **0.772** | **0.837** |
| Without content features | 0.814 | 0.631 | 0.711 |
| Without structural features | 0.618 | 0.576 | 0.596 |
| Without temporal features | 0.901 | 0.769 | 0.830 |
| Without history features | 0.887 | 0.782 | 0.831 |

为了研究各类特征对于模型预测能力的重要性,本文通过移除相应类别特征的方法,来考察其对模型预测性能的影响。实验结果表明,移除内容特征后,$F$值从83.7%下降到71.1%。因此,内容特征对于用户转发选择预测具有指示作用。在移除结构特征后,$F$值更是下降到59.6%。由此可见用户之间的结构特征,对于用户转发选择行为具有重要的指示作用。而在分别移除简单时序特征和交互历史特征后,模型预测性能并没有显著的下降,由此判断,这两类信息对于用户的转发选择行为的影响较小。

本文将模型训练所得参数列于表3中,从而可以更为清晰地看出各个特征所起的作用。例如特征11(表示Bob是否为消息的原发者)与Allen转发暴露源Jim的概率呈负相关,即当Bob为消息的原发者时,Allen更倾向于转发暴露源Bob而不是Jim。再比如特征8(表示Jim的入度是否大于Bob)与Allen转发暴露源Jim的概率呈正相关,即当Jim的入度大于Bob时,Allen更倾向于转发入度大的暴露源Jim而不是Bob。

### 5

本文以新浪微博为场景,围绕信息传播中的多次暴露现象展开研究,探索信息传播过程中多次暴

Table 3　Feature coefficients
表3

| 模型参数 | 特征编号 | 系数 |
|---|---|---|
| $\beta_1$ | 1 | −0.142 |
| $\beta_2$ | 2 | −0.014 |
| $\beta_3$ | 3 | −0.086 |
| $\beta_4$ | 4 | 0.042 |
| $\beta_5$ | 5 | 0.014 |
| $\beta_6$ | 6 | 0.004 |
| $\beta_7$ | 7 | 0.012 |
| $\beta_8$ | 8 | 0.213 |
| $\beta_9$ | 9 | 0.135 |
| $\beta_{10}$ | 10 | −0.106 |
| $\beta_{11}$ | 11 | −0.246 |
| $\beta_{12}$ | 12 | −0.068 |
| $\beta_{13}$ | 13 | 0.003 |
| $\beta_{14}$ | 14 | 0.017 |
| $\beta_{15}$ | 15 | −0.014 |
| $\beta_{16}$ | 16 | 0.057 |

露情形下用户转发行为的基本规律。针对用户在多个消息暴露源下的转发选择预测问题,融合了消息内容、暴露源的结构、时序以及交互历史等多方面因素,建模和预测了个体的转发选择行为。实验结果表明,在融合上述特征后,模型的预测准确率高达91.3%,其中结构特征和内容特征对于用户转发选择行为具有重要的指示作用。

本文后续研究方向包括量化用户之间话题层面的影响力和易受影响程度,探索和建模系统外部事件对用户转发行为的影响,并结合多次暴露对于用户转发行为所带来的累积效应,提出了一个用户转发行为预测的概率模型框架。此外,本文后续还将研究个体的转发选择行为与消息流行度及其动态过程之间的关联关系,利用消息早期的扩散信息,通过对微观个体行为的建模,来预测消息的未来流行度。为了进一步验证本文的泛化能力和适用场景,未来还将在多个语料集上进行方法的验证和扩展。

### References:

[1] Lazer D, Pentland A, Adamic L, et al. Computation social science[J]. Science, 2009, 323(5915): 721-724.

[2] Li Deyi, Zhang Tianlei, Huang Liwei. A down-to-earth cloud computing: location-based service[J]. Chinese Journal of Electronics, 2014, 42(4): 786-790.

[3] Barabási A-L. The origin of bursts and heavy tails in human dynamics[J]. Nature, 2005, 435: 207-211.

[4] Song Chaoming, Qu Zehui, Blumm N, et al. Limits of predictability in human mobility[J]. Science, 2010, 327(5968): 1018-1021.

[5] Katz E. The two-step flow of communication: an up-to-date report on a hypothesis[J]. Public Opinion Quarterly, 1957, 21(1): 61-78.

[6] Wu Shaomei, Hofman J M, Mason W A, et al. Who says what to whom on Twitter[C]//Proceedings of the 20th International Conference on World Wide Web, Hyderabad, India, Mar 28-Apr 1, 2011. New York: ACM, 2011: 705-714.

[7] Liben-Nowell D, Kleinberg J. Tracing information flow on a global scale using Internet chain-letter data[J]. Proceedings of the National Academy of Sciences of the United States of America, 2008, 105(12): 4633-4638.

[8] Leskovec J, Adamic L, Huberman B. The dynamics of viral marketing[J]. ACM Transactions on the Web, 2007, 1(1): 1-28.

[9] Yang J, Leskovec J. Patterns of temporal variation in online media[C]//Proceedings of the 4th ACM International Conference on Web Search and Data Mining, Hong Kong, China, Feb 9-12, 2011. New York: ACM, 2011: 177-186.

[10] Suh B, Hong Lichan, Pirolli P, et al. Want to be retweeted? Large scale analytics on factors impacting retweet in Twitter network[C]//Proceedings of the 2010 IEEE 2nd International Conference on Social Computing, Minneapolis, USA, Aug 20-22, 2010. Piscataway, USA: IEEE, 2010: 177-184.

[11] Romero D M, Meeder B, Kleinberg J. Differences in the mechanics of information diffusion across topics: idioms, political hashtags, and complex contagion on Twitter[C]//Proceedings of the 20th International Conference on World Wide Web, Hyderabad, India, Mar 28-Apr 1, 2011. New York: ACM, 2011: 695-704.

[12] Myers S A, Zhu Chenguang, Leskovec J. Information diffusion and external influence in networks[C]//Proceedings of the 18th ACM SIGKDD Conference on Knowledge Discovery and Data Mining, Beijing, Aug 12-16, 2012. New York: ACM, 2012: 33-41.

[13] Macskassy S A, Michelson M. Why do people Retweet? anti-homophily wins the day![C]//Proceedings of the 5th International Conference on Weblogs and Social Media, Barcelona, Spain, Jul 17-21, 2011. Palo Alto, USA: AAAI, 2011: 209-216.

[14] Aral S, Walker D. Identifying influential and susceptible members of social networks[J]. Science, 2012, 337(6092): 337-341.

[15] Bao Peng, Shen Huawei, Chen Wei, et al. Cumulative effect in information diffusion: empirical study on a microblogging network[J]. PLoS ONE, 2013, 8(10): e76027.

[16] Bao Peng, Shen Huawei, Huang Junming, et al. Popularity prediction in microblogging network: a cxase study on Sina-weibo[C]//Proceedings of the 22nd International Conference on World Wide Web, Rio de Janeiro, Brazil, Apr 7-11, 2013. New York: ACM, 2013: 177-178.

[17] Bao Peng, Shen Huawei, Jin Xiaolong, et al. Modeling and predicting popularity dynamics of microblogs using self-excited Hawkes processes[C]//Proceedings of the 24th International Conference on World Wide Web, Florence, Italy, Mar 18-22, 2015. New York: ACM, 2015: 9-10.

[18] Ugander J, Backstrom L, Marlow C, et al. Structural diversity in social contagion[J]. Proceedings of the National Academy of Sciences of the United States of America, 2012, 109(16): 5962-5966.

[19] Zhu Yuxiao, Zhang Xiaoguang, Sun Guiquan, et al. Influence of reciprocal links in social networks[J]. PLoS ONE, 2014, 9(7): e103007.

[20] Huang Junming, Li Chao, Wang Wenqiang, et al. Temporal scaling in information propagation[J]. Scientific Reports, 2014, 4: 5334.

[2] 李德毅, 张天雷, 黄立威. 位置服务: 接地气的云计算[J]. 电子学报, 2014, 42(4): 786-790.

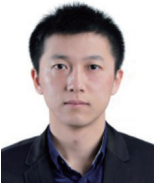

BAO Peng was born in 1987. He received the Ph.D. degree in computer science from Institute of Computing Technology, Chinese Academy of Sciences in 2015. Now he is an assistant professor and M.S. supervisor at Beijing Jiaotong University. His research interests include social media analytics, information propagation and network science.
鲍鹏(1987—),男,安徽六安人,2015年于中国科学院计算技术研究所获得博士学位,现为北京交通大学讲师、硕士生导师,主要研究领域为社会媒体分析,信息传播,网络科学。作为科研骨干参与了863课题、973课题和国家自然科学基金等重要科研任务。

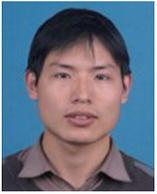

SHEN Huawei was born in 1982. He received the Ph.D. degree in computer science from Institute of Computing Technology, Chinese Academy of Sciences in 2010. Now he is an associate professor and M.S. supervisor at Institute of Computing Technology, Chinese Academy of Sciences. His research interests include social media analytics, information propagation and network science.
沈华伟(1982—),男,河南太康人,2010年于中国科学院计算技术研究所获得博士学位,现为中国科学院计算技术研究所副研究员、硕士生导师,主要研究领域为社会媒体分析,信息传播,网络科学。发表学术论文60余篇,主持国家自然科学基金项目3项,并承担863课题和973课题等重要科研任务。

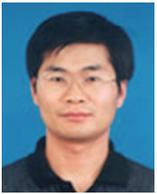

CHENG Xueqi was born in 1971. He received the Ph.D. degree in computer science from Institute of Computing Technology, Chinese Academy of Sciences in 2006. Now he is a professor and Ph.D. supervisor at Institute of Computing Technology, Chinese Academy of Sciences. His research interests include Web information retrieval, social media analytics and network data science.
程学旗(1971—),男,安徽安庆人,2006年于中国科学院计算技术研究所获得博士学位,现为中国科学院计算技术研究所研究员、博士生导师,中国科学院网络数据科学与技术重点实验室主任,主要研究领域为网络信息检索,社会媒体分析,网络数据科学。发表学术论文100余篇,主持10余项国家自然科学基金、973课题、863课题等重要科研项目,2014年获国家杰出青年科学基金资助。